\documentclass[prl, aps, notitlepage, superscriptaddress, showpacs, a4paper,twocolumn,10pt]{revtex4-1} 

\usepackage{graphicx}  
\usepackage{bm}        
\usepackage{amssymb}   
\usepackage[T1]{fontenc}
\usepackage[english]{babel}
\usepackage{subfigure}
\usepackage{epsfig}
\usepackage{verbatim}
\usepackage{setspace}
\usepackage{placeins}
\usepackage{amsmath}
\usepackage{mathrsfs}
\usepackage{slashed}
\usepackage[usenames, dvipsnames]{color}
\usepackage{graphics}
\usepackage[margin=1in,footskip=0.25in]{geometry} 
\usepackage[bookmarks=false,pdfborder={0 0 0}]{hyperref}
\usepackage[mediumspace,mediumqspace,squaren]{SIunits}
\usepackage{amsfonts}
\hypersetup{
colorlinks=true,
linkcolor=blue,
citecolor=blue}

\begin{document}

\title{Compensation of Beer-Lambert attenuation using \\ non-diffracting Bessel beams}

\author{Quentin Fontaine}
\affiliation{Laboratoire Kastler Brossel, Sorbonne Universit\'e, CNRS, ENS-PSL Research University, Coll\`ege de France, Paris 75005, France}

\author{Huiqin Hu}
\affiliation{Laboratoire Kastler Brossel, Sorbonne Universit\'e, CNRS, ENS-PSL Research University, Coll\`ege de France, Paris 75005, France}
\affiliation{State Key Laboratory of Precision Spectroscopy, East China Normal University, Shanghai, 200062, China}

\author{Simon Pigeon}
\affiliation{Laboratoire Kastler Brossel, Sorbonne Universit\'e, CNRS, ENS-PSL Research University, Coll\`ege de France, Paris 75005, France}

\author{Tom Bienaim\'e}
\affiliation{Laboratoire Kastler Brossel, Sorbonne Universit\'e, CNRS, ENS-PSL Research University, Coll\`ege de France, Paris 75005, France}

\author{E Wu}
\affiliation{State Key Laboratory of Precision Spectroscopy, East China Normal University, Shanghai, 200062, China}

\author{Elisabeth Giacobino}
\affiliation{Laboratoire Kastler Brossel, Sorbonne Universit\'e, CNRS, ENS-PSL Research University, Coll\`ege de France, Paris 75005, France}

\author{Alberto Bramati}
\affiliation{Laboratoire Kastler Brossel, Sorbonne Universit\'e, CNRS, ENS-PSL Research University, Coll\`ege de France, Paris 75005, France}

\author{Quentin Glorieux}
\email[Corresponding author: ]{quentin.glorieux@sorbonne-universite.fr}
\affiliation{Laboratoire Kastler Brossel, Sorbonne Universit\'e, CNRS, ENS-PSL Research University, Coll\`ege de France, Paris 75005, France}

\begin{abstract}
We report on a versatile method to compensate the linear attenuation in a medium, independently of its microscopic origin.
The method exploits diffraction-limited Bessel beams and tailored on-axis intensity profiles which are generated using a phase-only spatial light modulator.
This technique for compensating one of the most fundamental limiting processes in linear optics is shown to be efficient for a wide range of experimental conditions (modifying the refractive index and the attenuation coefficient).
Finally, we explain how this method can be advantageously exploited in applications ranging from bio-imaging light sheet microscopy to quantum memories for future quantum communication networks.
\end{abstract}


\maketitle

\section{Introduction}
Imaging through diffusive media is an incredibly difficult task for an optician.
It spans topics from imaging through atmospheric clouds, to in-vivo microscopy, to imaging in dense atomic gases.
Several techniques are currently intensively investigated including speckle correlations ~\cite{katz2014non} and transmission matrix reconstruction ~\cite{popoff2010measuring}.
In bio-imaging, light-sheet or selected plane illumination microscopy allows selective illumination of tissues and fast 3D imaging of live organisms at the cellular scale which are much more precise than conventional confocal or multi-photon imaging ~\cite{keller2008reconstruction,huisken2004optical}.
However, the main limitation for the field-of-view of light-sheet microscopy is the penetration depth of the illumination through the tissue ~\cite{fahrbach2010microscopy}.
This fundamental limitation makes light-sheet imaging of deep tissues in living animals a challenging task, as it is complicated to illuminate deep structures effectively.
In this paper we propose and implement a general method to resolve this strong limitation.
We design a non-diffracting Bessel beam with the intensity in the central spot exponentially rising as function of the propagation, in order to exactly compensate for the losses in the tissue (or in any lossy medium).

Actually, the exponential attenuation of the beam due to scattering is common in optics.
As soon as a beam propagates through a medium, scattering will inevitably degrade the signal.
Our technique is of major interest not only for imaging through biological samples but it could also be used to study disordered media or light propagation in dilute atomic clouds and non-linear crystals.
In all these systems the exponential attenuation of a  beam due to linear losses is a fundamental limitation ~\cite{mccormick2008strong,glorieux2010double,jasperse2011relative,glorieux2012generation}.  

Introduced by Durnin \emph{et al} in 1987 ~\cite{Durnin:1987}, zero-order Bessel beams are one of the most representative "non-diffracting" solution of the Helmholtz equation with Airy (or parabolic) beams ~\cite{Bandres:2004}.
These optical fields result from the interference of an infinite number of plane waves whose wave-vectors constitute the generating lines of the so-called Bessel cone. 
The radial intensity profile of zero-order Bessel beams is described by the zero-order Bessel function of the first kind; a high intensity central peak is surrounded by an infinity of concentric rings of decreasing intensity.
Although perfect Bessel beams are only mathematical objects as they should carry an infinite energy, spatially limited quasi-Bessel beams can be realized experimentally.
Those beams have found various applications -- in optical trapping ~\cite{Cizmar:2005, Chavez:2002}, laser machining ~\cite{Courvoisier:2013}, nonlinear optics \cite{Johannisson:2003, Porras:2004, Polesana:2008} and imaging ~\cite{Dufour:2006, Zhao:2014} for example -- as their central cores  stay collimated on a distance that is orders of magnitude longer than the Rayleigh length.  

The modification of the intensity profile in the propagation direction has been recently studied experimentally in the context of counterbalancing the intensity decay induced by light absorption in weakly absorbing dye solutions \cite{Dorrah:2016}.
The general idea is to tailor the on-axis intensity of a Bessel beam.
It has been reported the use of an exicon (exponential intensity axicon) \cite{Golub:2012,Golub:09} and the generation of attenuation-resistant beams with computed generated holograms ~\cite{Dorrah:2016,Zamboni-Rached:2006}.
This last approach is known as frozen waves and results from the superposition of equal frequency Bessel beams produced by modulating only the amplitude of an incident plane-wave with a Spatial Light Modulator (SLM) ~\cite{Zamboni-Rached:04,Zanarella:18,PhysRevA.92.043839,Vieira:12,Vieira:14}.  

In this paper, we report on a more general and versatile method based on both phase and amplitude shaping of an incident Gaussian beam.

It allows for compensating any absorption coefficients up to $\alpha \! = \! 200$~m$^{-1}$, independently of the refractive index and of loss mechanism by using real space shaping with a reflective phase-only SLM.

We demonstrate experimentally the accuracy of this method in two very different media: a scattering sample based on a dilute solution of milk in water (index $n=1.33$) and an absorptive sample of near resonance atomic vapor ($n\simeq 1$).

In the first section, we present the theoretical background of the method including a general approach to compensate for the refractive index of the medium. 
A detailed description of our experimental setup follows, as well as the measurement of the on-axis intensity profile of the central peak in air.
In the next part, we present our results on the compensated absorption for two different media and show that our procedure is efficient independently of the refractive index and for a wide range of loss coefficients.
Finally, we describe the potential improvement of 2 orders of magnitude on the field of view for light-sheet microscopy using this approach.

\section{Shaping Bessel beams on-axis intensity}

At a given position $z$ on the optical axis (we assume $z  =  0$ in the following), the electric field in the transverse plane $E(r,z=0)$ of a radially symmetric laser beam can be expressed, under the paraxial and the scalar approximation, as an infinite superposition of zero order Bessel functions of the first kind $J_0$:
\begin{equation}
\label{ElectricField}
    E(r,z=0) = \frac{1}{2 \pi} \int_{0}^{\infty} S(k_{\perp},z=0) J_0(r k_{\perp}) k_{\perp} \mathrm{d} k_{\perp},
\end{equation}
\noindent where $r$ and $k_{\perp}$ stand respectively for the transverse radial coordinate and the associated spatial angular frequency. 
The spatial spectrum $S(k_{\perp},z=0)$ is the Hankel transform of the electric field amplitude $E(r,z=0)$.
The on-axis electric field $E(r=0,z)$ is obtained by taking the inverse Fourier transform of Eq.~\eqref{ElectricField} ~\cite{Cizmar:09}: 
\begin{multline}
   \label{OnAxisField}
    E(r=0,z) = \frac{1}{\pi} \int_{0}^{\infty} S(\sqrt{k_{0}^2 - k_{z}^2},z=0) \\
    \times \exp{\left( i k_{z} z\right)} \, k_{z} \, \mathrm{d} k_{z}.    
\end{multline}
where $k_{0} = 2\pi/\lambda$ is the laser wave-vector ($\lambda$ its wavelength) and $k_{z}  =  \sqrt{k_{0}^{2}-k_{\perp}^{2}}$ the longitudinal spatial frequency of a given Bessel mode. 
This formula give a physical insight about the engineering process we use to overcome attenuation. 
Each of the Bessel mode coming in the spectral decomposition Eq.~\eqref{ElectricField} will propagate in free-space with a slightly different longitudinal wave-vector $k_{z}$ and thus merge with different cone angles at distinct position along the optical axis.
The on-axis electric field results then from the interference arising between the individual modes. 
At the end of the day, if one wants to design a Bessel beam with a given on-axis intensity profile $I(z) = \vert E(r=0,z) \vert^{2}$ along the optical axis, the spatial spectrum $S$ must be engineered according to the following formula :
\begin{equation}
\label{Spectrum}
     S(k_{\perp}, z=0) \! = \! \frac{1}{k_{z}} \int_{0}^{\infty} \! \sqrt{I(z)} \, \exp{\left[i(k_{z0}-k_{z})z\right]} \, \mathrm{d}z,
\end{equation}
The spectrum $S$ is centered around the axial wave-vector of the target Bessel beam $k_{z0} = k_{0} \cos(\theta_{0})$.
The cone angle $\theta_{0}$ sets the spot size (the full width at half-maximum (FWHM) of the central peak in the transverse intensity profile), which is equal to $2.27/\left(k_{0} \sin (\theta_{0})\right)$ for a perfect zero order Bessel beam.
For non--evanescent modes, $k_{z}$ should lie in the interval $\left[k_{\min} \, , \,k_{0}\right]$, where $k_{\min}$ is defined by the numerical aperture (NA) of the imaging system as $k_{\min} = k_{0} \sqrt{1-\mathrm{NA}^{2}}$. 
As explained in Ref. ~\cite{Ouadghiri-Idrissi:16}, the target intensity profile should not vary on a length scale smaller than $\Delta z = 4 \pi / \left(k_{0}-k_{\min}\right)$ to avoid significant frequency truncations in the associated spectrum, leading to undesirable oscillations in the measured on--axis intensity profile.

The initial electric field $E(r,z=0)$ that will produce a Bessel beam with a given cone angle $\theta_{0}$ and an on-axis intensity profile $I(z)$ can be evaluated using Eq.~\eqref{ElectricField} and Eq.~\eqref{Spectrum}. 
In the following, we show how to generate the target beam by real-space shaping of an incident Gaussian beam on a Spatial Light Modulator (SLM). 
Fourier space shaping may also be considered ~\cite{Cizmar:09}. However, as the intensity distribution of a Bessel beam in Fourier space is a thin ring, the small overlap between the latter and the incident Gaussian profile will filter out most of the incident energy.
Higher efficiency can then be obtained using real space shaping. \\

We define $z=0$ to be the SLM plane position along the optical axis. 
Discretizing the electric field accordingly to the SLM matrix ($N_x\times N_y$), the target electric field $E(i,j,z=0^{+})$ right after the SLM can be decomposed in amplitude $A(i,j)$ and phase $\Phi(i,j)$ (where $0  \leq  i  \leq  N_{x}$ and $0  \leq  j  \leq  N_y$ stand for the pixel coordinates). 
As suggested by Davis \emph{et al.} ~\cite{Davis:99}, locally reducing the phase wrapping contrast allows for a modulation of the amount of light scattered in the first diffraction order, using a single hologram. 
We apply this technique with a phase-only SLM.
The expression of the SLM phase mask $\Psi$ is given by ~\cite{Bolduc:13, Ouadghiri-Idrissi:16}: 
\begin{equation}
\label{PhaseMask}
     \Psi(i,j) = M(i,j) \, \mathrm{mod} \left[ F(i,j) + \Phi_{\mathrm{g}}(i,j), \, 2 \pi \right].
\end{equation}
The function $F$ contains the phase information of the target electric field and $\Phi_{\mathrm{g}}$ stands for the grating linear phase ramp, used to separate the different diffraction orders in Fourier space. 
The modulo operation provides the phase wrapping. 
The diffraction efficiency is locally tuned by the modulation function $M$ ($0 \leq M(i,j) \leq 1$). 
The complex amplitude of the field diffracted in the first order can be expressed as follow ~\cite{Bolduc:13, Ouadghiri-Idrissi:16}:   
\begin{multline}
\label{FirstDiffraction}
    E_{1}(i,j,z=0^{+})  = A_{\mathrm{in}}(i,j) \, \mathrm{sinc} \left(\pi M(i,j) -\pi \right) \\ \times \exp{\left[ i \left( F(i,j) + \pi M(i,j) \right) \right]}, 
\end{multline}
where $A_{\mathrm{in}}$ is the amplitude of the incident laser beam on the SLM. 
By identifying $E_{1}$ with the target electric field, one can obtain the functions $F$ and $M$ solving the following system:  
\begin{align}
\label{System}
        M(i,j)  &= 1 + \frac{1}{\pi} \, \mathrm{sinc}^{-1} \left( \frac{A(i,j)}{A_{\mathrm{in}}(i,j)} \right) \\ 
        F(i,j)  &= \Phi(i,j) - \pi M(i,j) 
\end{align}
The inverse sinc function ($\mathrm{sinc}^{-1}$) is defined on $[-\pi,0]$. 
Computing it for each points of the hologram is usually demanding ($N_{x}  \times  N_{y}$ operations). 
However, if both the incident and the first order diffracted beams are radially symmetric, we only need to determine the radial profile of the modulation function. For beams centered in the SLM matrix, Eq.~\eqref{System} can be simplified such as: 
\begin{equation}
\label{Eq1System}
m(i) = 1 + \frac{1}{\pi} \, \mathrm{sinc}^{-1} \left( \frac{A(i,N_y/2)}{A_{\mathrm{in}}(i,N_y/2)} \right) ,
\end{equation}
where $i$ is an integer running from 0 to $N_{x}/2$ (for $N_{x} \geq N_{y}$).
Using a circular interpolation, $M$ can be entirely reconstructed from $m$, computing the inverse $\mathrm{sinc}$ function for $N_{x}/2$ points only instead of $N_{x}\times N_{y}$.
In practice, we start with the clean-up of the incident laser beam, filtering out in Fourier space its high k-vector components with a small pinhole aperture. 
Afterwards, the widths of the input Gaussian beam is radially symmetric in the SLM plane ($\omega_{x,y} \! \simeq \! 3.3 \pm 0.1$ mm). \\
In principle, arbitrary on-axis intensity profiles can be generated using the method described above. 
In the following section, we introduce the target profile $I(z)$ we use to maintain the central peak intensity constant along the propagation in an uniform and linear lossy medium. 
Our approach is independent of the loss origin; we demonstrate its validity for both absorbing and scattering type of losses.  
Let $L$ and $\alpha$ stand respectively for the propagation length and the linear attenuation coefficient of the medium.
According to the Beer-Lambert's law, the transmittance $t$ of the medium decays exponentially with the propagation distance: $T = \exp(-\alpha z)$. 
Therefore, to compensate for these losses, the on-axis intensity should exponentially increase along the propagation such as $I(z)\sim \exp(\alpha z)$.
We ramp the on-axis intensity up (from 0 to $I(z_1)=I_{0}$), until the entrance plane position $z_1$, before exponentially increasing it over the length $L$. 
We then make it go back to 0 smoothly. 
The full on-axis target profile we designed is described as:  
\begin{align}
\label{SystemProfile}
I(z) = 
    \begin{cases}
        I_{0}  \left( \frac{\sin(C_{1} z/z_{1})}{\sin(C_{1})} \right) ^{2} &\mathrm{if} \;\; 0 \leq z \leq z_{1} \\
        I_{0} \exp \left[ \alpha  (z-z_{1}) \right] &\mathrm{if} \;\; z_{1} \leq z \leq z_{2} \\
        I_{\max} \sin^{2} \left[ C_{2} + (\frac{\pi}{2} - C_{2})  \frac{z-z_{2}}{z_{3}-z_{2}} \right] \; &\mathrm{if} \;\; z_{2} \leq z \leq z_{3} \\
        I_{\max}\sin^{2} \left[\frac{\pi}{2} \left(1- \frac{z-z_{3}}{z_{4}-z_{3}} \right) \right]  &\mathrm{if} \;\; z_{3} \leq z \leq z_{4} .
    \end{cases}
\end{align}
For all the measurements we performed, we set $z_{1}  \times  G^2 = 1.5$ cm, $z_{2}  = z_{1} + \frac{L}{G^2}$ (with $L = 7.5$ cm) and $z_{4} = 3 \, z_{1} + \frac{L}{G^2}$, where $G = 0.5$ stands for the telescope demagnification factor which optically conjugates the SLM and the $z=0$ planes. 
$C_{1,2}$ and $z_{3}$ are constants chosen in order to make the profile continuous and differentiable; they are defined in the supplementary materials.
In the following, the profile has been normalized to 1 dividing $I(z)$ by the maximum intensity $I_{\max} = I_{0} \left( 1 + \exp(\alpha L) \right)$.
The spatial spectrum associated to this on-axis profile is analytically derived in the appendix.
We obtain the target electric field by computing the inverse Hankel transform using Eq.~\eqref{ElectricField}.\\

\noindent When the linear refractive index $n$ of the medium is not equal to 1 (as implicitly assumed before), the target Bessel beam will undergo refraction at both the entrance and the output plane of the medium. 
Applying the Snell's law at the entrance (resp. the output) plane, we get: $\sin(\theta_{i}) = n \sin(\theta_{r})$, where $\theta_{i}$ and $\theta_{r}$ stand respectively for the incident and refractive cone angle of a given Bessel mode.
Using the transverse spatial angular frequency $k_{\perp} = n  k_{0}  \sin(\theta)$, we find that $k_{\perp}^{(i)} = k_{\perp}^{(r)}$. 
Therefore, according to Eq.~\eqref{ElectricField}, the transverse shape of the target Bessel beam is not modified by successive refractions ~\cite{Mugnai:2009}. 
Nevertheless, the cone angle is modified when the Bessel beam enters the medium.

\begin{figure}[htbp]
\centering
\includegraphics[width=\linewidth]{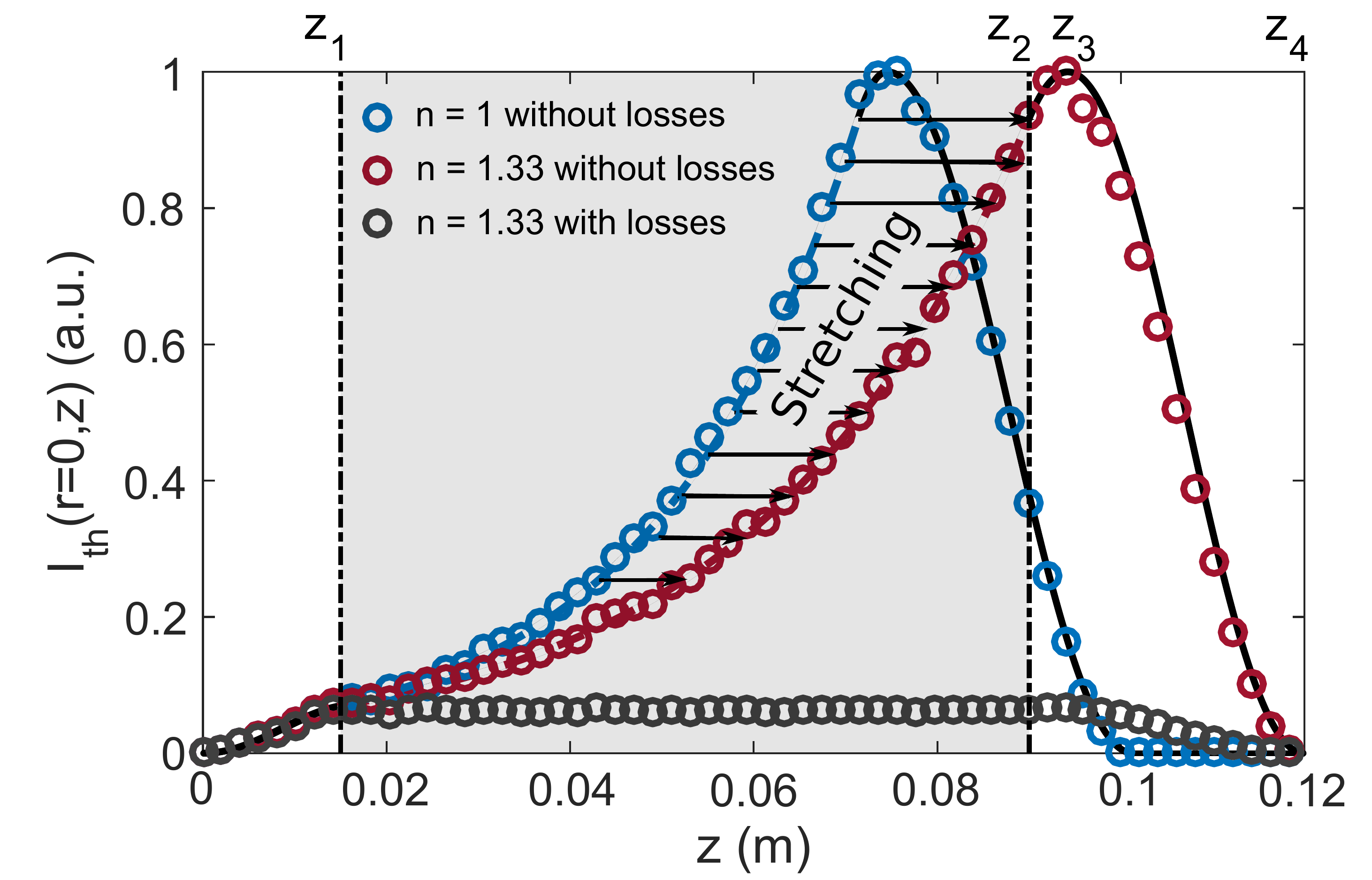}
\caption{Numerical simulation of the longitudinal refractive stretching of the on-axis profile. 
Blue, red and grey circles are obtained by solving numerically the evolution of the Bessel beam in air (blue), in a non-lossy (red) and in a lossy material of refractive index $n=1.33$. 
The simulation data obtained with the refractive medium can be deduced from the vacuum ones by stretching the z axis by a factor $n$ between $z_{1}$ and $z_{2}$. 
We adjust the exponentially growing section of the profile (blue dotted line) such that the stretched Bessel beam ends up compensating the attenuation. 
When losses are added, the on-axis intensity remains constant along the propagation (black dots) as expected.}
\label{fig:Stretching}
\end{figure}

As $n > 1$, the inner cone angle $\theta_{r}$ is lower than the external one and the Bessel beam will lengthen more than in air. 
This stretching of the beam inside the medium will necessarily reduce the compensation coefficient by a factor $n$. 
To counteract this effect, we constrict the exponentially rising part of the target on-axis profile beforehand by a factor $n$ (as suggested in ~\cite{Golub:12}). 
In other words, we replace $L$ by $L/n$ and $\alpha$ by $\alpha n$ in the second line of Eq.~\eqref{SystemProfile}. 
By doing so, the stretching of the beam will be controlled, and so compensate exactly for the exponential attenuation in the medium, as shown in Fig. \ref{fig:Stretching}.
This compensation procedure generalizes easily to more complex situations when several layers of materials (with different attenuation coefficients and refractive indices) are involved. 

\section{Experiment}
\label{sec:examples}

\subsection{Experimental setup}

\begin{figure}[htbp]
\centering
\includegraphics[width=\linewidth]{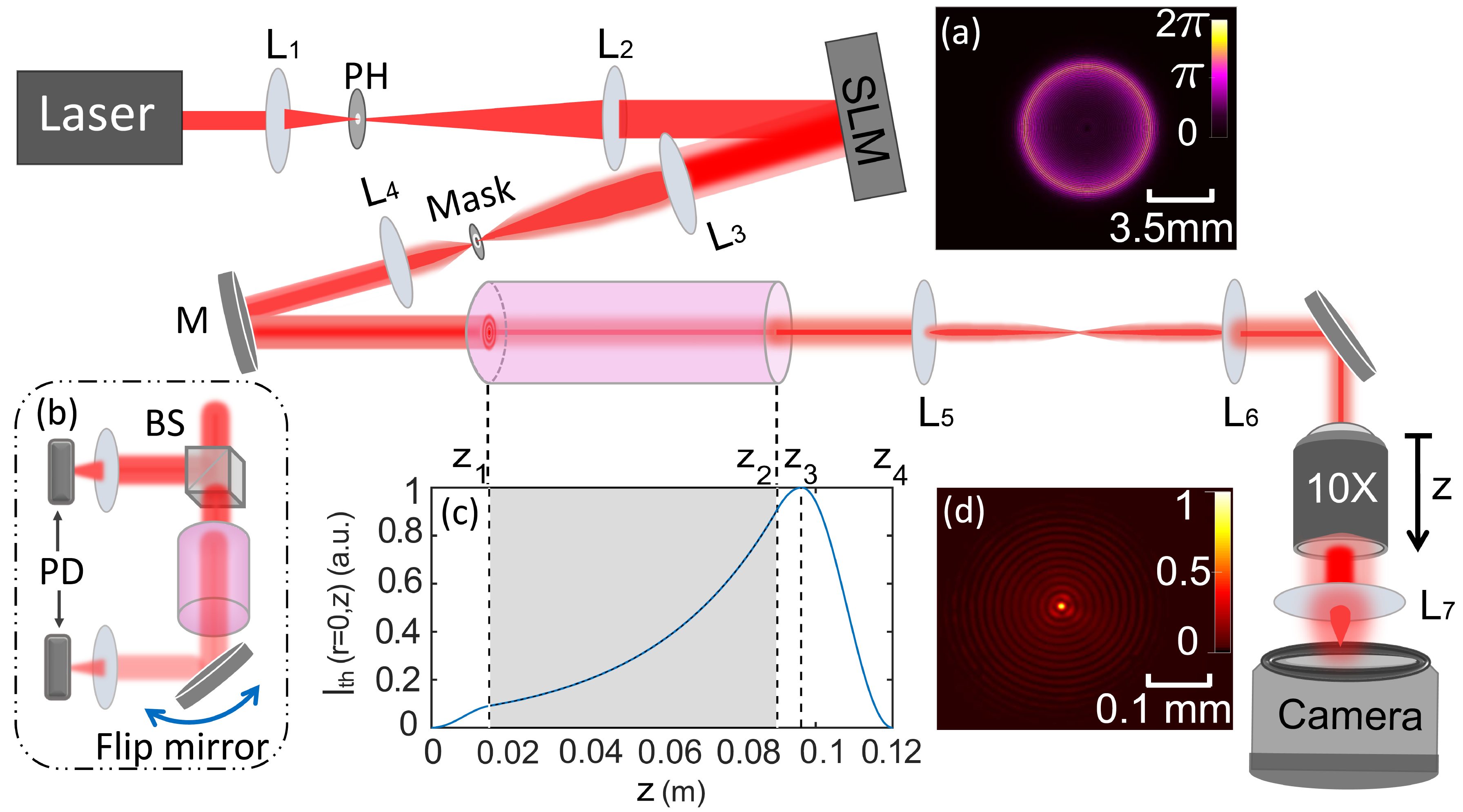}
\caption{Experimental setup. 
$L_{1-7}$ label the different lenses. 
PH is a pinhole used to clean up the beam. 
An iris and a mask (thin metallic dot on a glass window) cut all the diffraction orders (except the first one) in the Fourier space of $L_{3}$. 
The mirror $M$ sets on a translation stage to adjust the $L_{4}$ focal plane position, where the Bessel beam start forming.
Insets:
(a) Phase mask applied on the SLM.
No grating was added on top. 
(b) Transmission measurement setup. 
A wide (non-saturating) Gaussian beam splits on a (10:90) beamsplitter; the most intense part propagates through the lossy medium. 
The beams are focused on two photodiodes (PD) in order to monitor both the stability of the laser intensity and the material transmission. 
(c) Target on-axis intensity profile. 
The shadowed region shows where the lossy material should be positioned. 
(d) Transverse profile of the generated Bessel beam.}
\label{fig:ExperimentalSetup}
\end{figure}

\noindent Our experimental setup is shown in Fig. \ref{fig:ExperimentalSetup}.
A continuous-wave laser beam, produced by a tapered amplifier laser system, gets 4 times magnified by a 4-f telescope system (lenses $L_{1}$ and $L_{2}$) and is spatially filtered in the Fourier space of $L_1$.
The resulting $6.6\,$mm diameter radially symmetric Gaussian beam reaches the center of the SLM chip with a normal incidence.
The SLM used for the experiment is a liquid crystal on silicon (LCOS) phase-only modulator, with an effective area of $1272 \times 1024$ pixels and a pitch of 12.5 $\micro$m. 
After shaping, a $50:50$ non-polarizing beam splitter separates the diffracted beam from the incoming one. 
Another 4-f telescope ($L_{3}$ and $L_{4}$, $\mathrm{NA} = 0.017$) optically conjugates the SLM and the $z=0$ planes with a demagnification factor $G=0.5$. 
The choice of the telescope lenses $L_{3}$ and $L_{4}$, as well as the demagnification factor $G$, is conditioned by the length of the lossy medium we are dealing with.
For biological applications, $G$ should be divided by 10 at least, as pointed out in section \ref{sec:Result}. 
The first order diffracted beam is then selected by masking the zero and the higher order ones in the Fourier plane of $L_{3}$.
The Bessel beam finally starts forming from the focal plane of $L_{4}$ (at $z=0$) and propagates through a lossy medium.
The output plane of the medium is imaged by a third 4-f arrangement (lenses $L_{5}$ and $L_{6}$, $\mathrm{NA}'= 0.042$) onto a microscope objective which is set up on a computer controlled translation stage. 
By moving the objective along the optical axis, we can monitor the Bessel beam evolution along $z$. 
The last lens ($L_{7}$) images the plane we look at on the CMOS camera. 
The magnification factor $G'$ of the whole imaging system is $13.6 \pm 0.1$.    

\subsection{Results and discussion}
\label{sec:Result}
\begin{figure}[htbp]
\centering
\includegraphics[width=\linewidth]{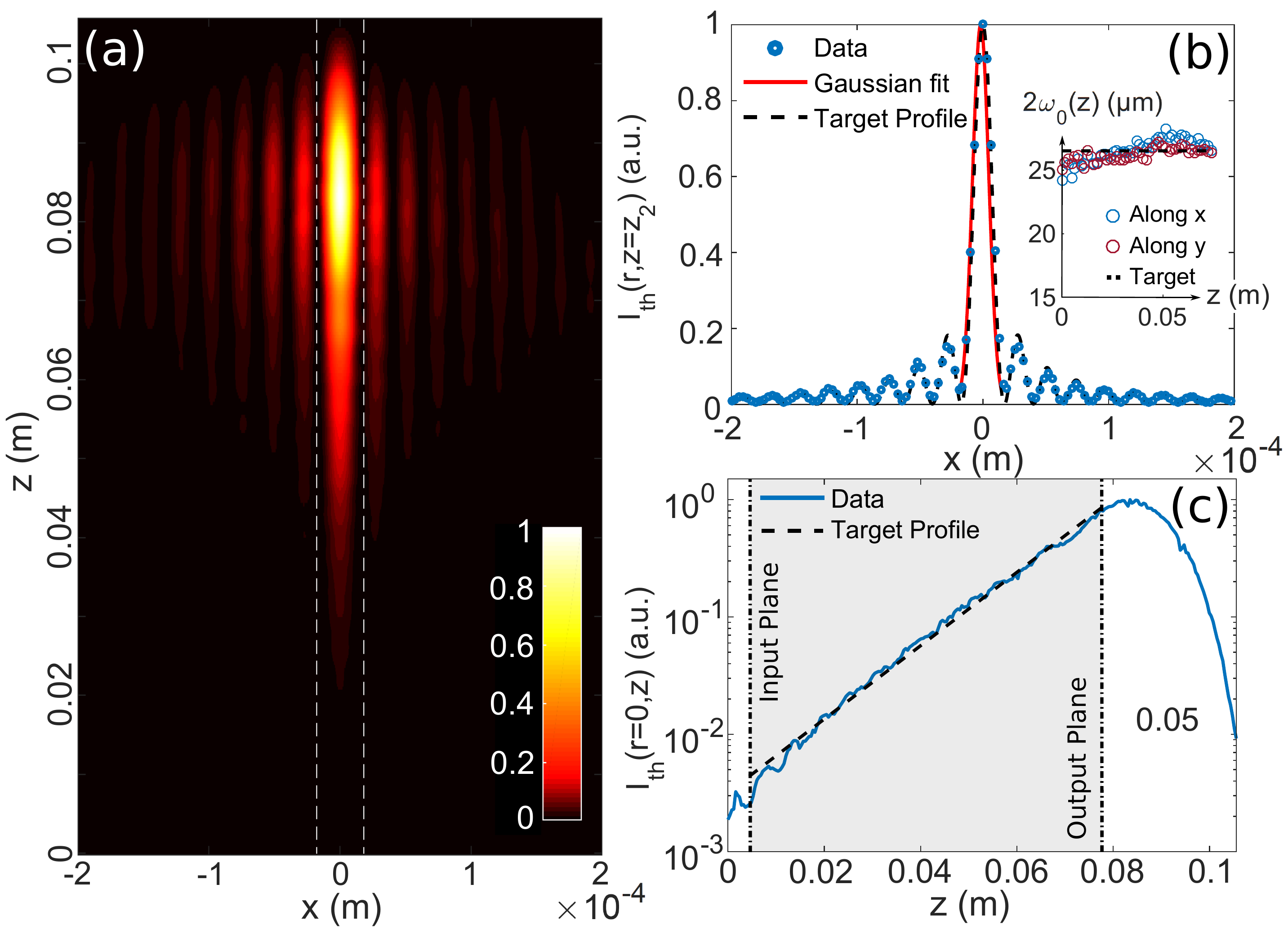}
\caption{Experimental characterization of the reconstructed Bessel beam.
The Bessel cone angle $\theta_0$ was set to $(1/G) \times 8.5$ mrad.
The 2D map fig. (a) is obtained by scanning slowly ($v = 2$ mm.s$^{-1}$) the microscope objective along the z axis. 
The white dotted lines on both sides of the central peak define the region where the Gaussian fit is performed.
The blue curves on fig. (b) and (c) are obtained cutting the 2D map along $z = 0.083$ (at the maximum on axis intensity position) and $x=0$ respectively.}
\label{fig:ProfileAll}
\end{figure}

The first step to verify the compensation of the beam attenuation is to compare the transverse and longitudinal intensity profiles of the experimentally measured beam with the target ones from simulations, in air. 
As shown in Fig. \ref{fig:ProfileAll}, we design the Bessel beam to overcome $96\%$ attenuation over a lossy, $7.5$ cm long medium.
The 2D map in Fig. \ref{fig:ProfileAll} (a) is obtained by scanning slowly ($v = 2$ mm.s$^{-1}$) the microscope objective along the z axis. 
Both the transverse, Fig. \ref{fig:ProfileAll} (b), and the longitudinal, Fig. \ref{fig:ProfileAll} (c), measured intensity distributions of the tailored beam (blue circle and line) are in excellent agreement with the simulation (dashed black lines). 
The target profiles (dashed black lines) are obtained by solving numerically the evolution of the transverse electric field from $z=0$ to $L$ with the second order split-step method. 
We take as initial condition a field with the SLM imprinted phase $\Psi$ and the radially symmetric Gaussian envelope of the SLM input beam.
To determine accurately the central peak intensity along $z$ as presented in Fig. \ref{fig:ProfileAll} (c), we fit with a Gaussian profile the region delimited by the two white dashed lines on both sides of the central peak as illustrated in Fig. \ref{fig:ProfileAll} (b) (red line). 
The width of the peak along the propagation is found to be constant ($\pm 5\%$), as shown in the inset of Fig. \ref{fig:ProfileAll} (b); we are therefore able to control the longitudinal intensity profile without altering the non-diffracting behavior of the Bessel beam. 
Nevertheless, small amplitude oscillations can be observed at the beginning of the measured on-axis profile Fig. \ref{fig:ProfileAll} (b). 
They are due to high longitudinal frequency truncation, as $k_{z}$ is upper bounded by the laser wave-vector $k_{0}$.
We can reduce the oscillation amplitude by increasing the Bessel cone angle $\theta_{0}$.
In our setup we are limited by the mirror size as the beam will start to clip and will loose its radial symmetry.


\noindent The lossy medium is then positioned on the beam path. 
By fitting the on-axis intensity profile with the function of Eq.~\eqref{SystemProfile}, we find the position $z_{2}$ where the medium output plane should set. 
We then move the lossy medium cell along the optical axis until imaging this plane on the camera. 
The $1$ mm depth-of-field of the imaging system and the standard deviation on the fit parameters translate into an uncertainty of $\pm 2$~mm on the medium output plane position. \\

\noindent Three different media (contained in three different glass cells) have been used to check our ability to compensate for the attenuation of the Bessel peak intensity along propagation. 
Two cells are filled with isotopically pure Rubidium vapor (the first ($7.5$ cm long) with $^{87}$Rb only and the second ($2.5$ cm long) with $^{85}$Rb only); the third one ($2.5$ cm long) contains a diffusive water-milk mixture. 
Rubidium cells are heated up to 140$\degree$C. At this temperature, the atomic density is large ($n_{a} \simeq 2-5 \times 10^{13}$ atoms/cm$^{3}$).
By tuning the laser frequency $\nu_{0}$ over the $D_{2}$ Rubidium absorption lines, we can change the transmission over several orders of magnitude, without affecting significantly the refractive index $n_{\mathrm{Rb}}$. 
The latter is estimated theoretically taking the Rubidium hyperfine structure and the Doppler broadening into account ~\cite{Siddons:2008}: $n_{\mathrm{Rb}}(\nu_{0})  \simeq 1.00 \pm 0.02$ (scanning $\nu_{0}$ over the whole absorption spectrum). 
The transmission of the water-milk mixture can be tuned changing the milk concentration. 
Remaining under highly diluted condition, the medium refractive index stays close to the water one $n_w \simeq 1.33$. 
As explained above, we should balance in this case the change of refracting index stretching the Bessel beam along the optical axis, replacing beforehand in the target profile $L$ and $\alpha$ with $L/n$ and $\alpha \, n$ respectively. 

\begin{figure}[htbp]
\centering
\includegraphics[width=\linewidth]{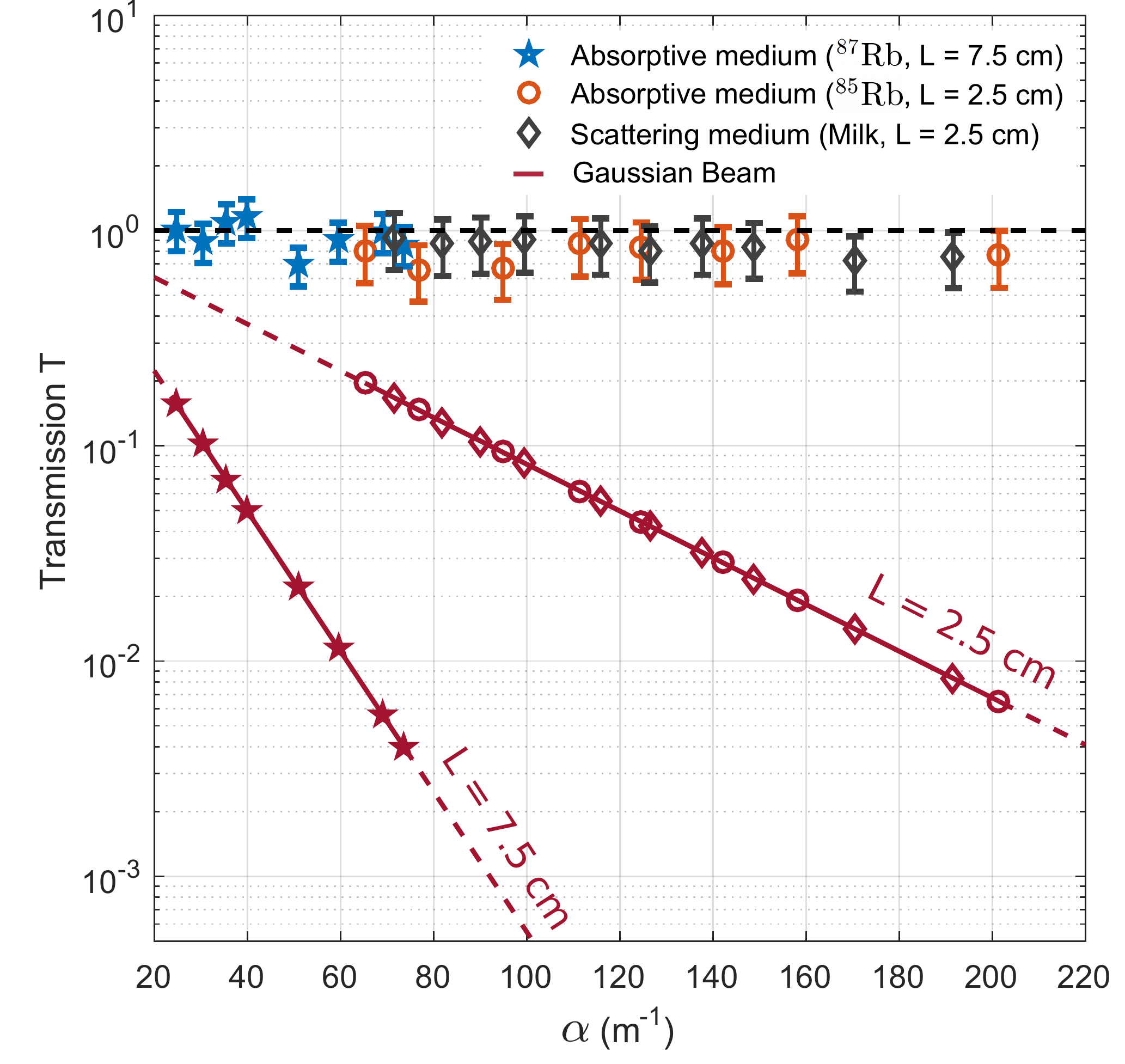}
\caption{Transmission T as a function of the attenuation coefficient $\alpha$. 
The transmission of the Bessel beam through the Rubidium vapors are plotted in blue stars ($^{87}$Rb) and orange circles ($^{85}$Rb). Data obtained with the water-milk mixture are plotted in grey diamonds. 
The two red lines show the transmission expected from the Beer-Lambert law for 2.5 cm and 7.5 cm long lossy materials.}
\label{fig:Transmission}
\end{figure}

We design the target profile to overcome attenuation over 7.5 cm long materials, whatever the length of the cell we use. 
The overall Bessel power is reduced to keep the input peak intensity lower than the Rubidium saturation intensity ($I_{\mathrm{sat}} \simeq  2.5$ mW/cm$^{2}$ for linearly polarized laser beam).
We finally measure both the peak intensity in the entrance plane (without cell) and in the output plane (with cell) to evaluate the transmission trough the medium. We perform the fitting procedure on five different images of the central spot with a 2D-Gaussian function as detailed before. 
The measured transmission is shown in Fig. \ref{fig:Transmission}. 
The experimental data obtained with the $^{87}$Rb$\,$vapor cell, the $^{85}$Rb one and the water-milk mixture are respectively plotted in blue stars, orange circles and grey diamonds. 
The $4 \%$ reflectively of the cell windows has been taken into account. 
The black dashed line represents a perfect on-axis compensation ($T=1$); most of the experimental points lie right under it. The small discrepancy comes from the input plane intensity measurements rather than from the output plane ones.   
The oscillations of the intensity at the medium input plane, visible on Fig. \ref{fig:ProfileAll} (c), adding to a positioning uncertainty of $4$ mm of the input plane, lead to the errorbar uncertainty reported in Fig. \ref{fig:Transmission}.   
The red lines represent the transmission of a non-saturating collimated Gaussian beam with respect to the attenuation coefficient $\alpha$ for a 2.5 cm and a 7.5 cm long lossy medium. \\

\section{Applications}
\noindent Compared to previous experiments, in which compensation for $30 \, \%$ and $10 \, \%$ attenuation have been achieved using exponential intensity axicon ~\cite{Golub:12} and attenuation-resistant frozen waves ~\cite{Dorrah:16} respectively, we manage to maintain the on-axis intensity of the Bessel beam quasi-constant along its propagation in media with an attenuation up to $99.995 \%$. 
This is a crucial advantage for biological applications as the diffusive coefficients of biological tissues observed in light-sheet microscopy range typically from $50$ to $200$ cm$^{-1}$ ~\cite{Johns:05, Nylk:2018}.
For comparison, let's assume that the sample transmission is $6 \times 10^{-3}$ under Gaussian illumination (as for the last orange circle of Fig. \ref{fig:Transmission}) for a diffusive coefficient of $200$ cm$^{-1}$. 
The length of the sample is then equal$\,$to $ -\ln{(6 \times 10^{-3})}/200  \simeq  250 \, \micro$m. By increasing the demagnification of the 4-f telescope $\{L_{3}, L_{4}\}$ by 10, we could easily decrease the length of the Bessel zone by a factor 100 (as the length varies with $G^{2}$) and, using the same target profile, uniformly illuminate biological tissues over hundred of microns up to diffusive coefficient $\alpha$ of $200$ cm$^{-1}$. 
For such attenuation coefficient, the field of view would then be more than 100 $\micro$m longer than the best one obtained in the literature so far ~\cite{Nylk:2018} (or even more if partial attenuation-compensation is considered).\\

On the other hand, any light-matter interface that relies on maximising  interactions over a long distance along the propagation axis, such as EIT-based quantum memory or gradient echo memory ~\cite{glorieux2012temporally}, will benefit from this technique to improve significantly the effective optical depth.
As it is known that the quantum memory efficiency (for Raman schemes) is proportional to the Rabi frequency of the coupling field ~\cite{hetet2008photon}, compensating for losses will immediately increase the storage efficiency especially in the case of an ultra-high optical depth medium ~\cite{sparkes2013gradient}.
\noindent Finally, attenuation compensated Bessel beams are needed for the generation of stationary non-diffracting potentials in fluid of light experiments in the propagation configuration ~\cite{Carusotto:2014}, where superfluidity has  recently been observed ~\cite{Michel:2018,Fontaine:18}. 
So far, exponential attenuation of the defect was the main limitation of these experiments in hot atomic media.
With the approach described in this paper, future experiments can be envisioned where a Bessel beam pumps a vapor close to resonance and modifies the medium refractive index locally in the transverse plane and uniformly along the beam axis. 

\section{Conclusions}
We have reported the shaping of the longitudinal intensity profile of Bessel beams using a phase-only SLM.
We have shown that this method can be used to compensate the Beer-Lambert law and generate a constant intensity profile along the propagation direction in a lossy medium. 
We verified that our approach is robust independently of the loss mechanism at play in the medium and for a wide range of conditions (various refraction indices and attenuation coefficients). 
The results are in agreement with numerical simulations and the most crucial limitations are clearly identified. 
We identified two applications where this can be advantageously used: in bio-imaging and in quantum optics. 
Finally, this method can be easily generalized to tailor any kind of on-axis intensity profile.

\section*{Funding Information}

This work has been supported by the C-FLigHT ANR project and PhoQus Quantum Flagship and by ECNU scholarship program for graduate students. H.H. is supported by the Caiyuanpei Programm. Q.G. and A.B. are members of the Institut Universitaire de France (IUF).

\section*{Supplementary materials}

\subsection{On-axis intensity profile and spatial spectrum}

The target profile we designed to compensate on the optical axis the attenuation of the central peak intensity is given by Eq.~\eqref{SystemProfile}. The SLM mask is optically conjugated with the $z=0$ plane by the 4-f telescope formed with the lenses $L_3$ and $L_4$. For clarity, we have ignored the telescope magnification factor $G = 0.5$ in the analytical derivations above, but we need to consider it in practice when we compute the target intensity profile. For all the measurements we performed, we set $z_{1}  \times  G^2 = 1.5$ cm, $z_{2}  = z_{1} + \frac{L}{G^2}$ (with $L = 7.5$ cm) and $z_{4} = 3 \, z_{1} + \frac{L}{G^2}$. We choose $C_{1}$ and $C_{2}$ in order to make the target profile continuous and differentiable at $z_{1}$ and $z_{2}$: the constant $C_{1}$ is obtained by solving the equation $\tan(x) = \frac{2 x}{\alpha L}$ (deriving from the diffentiability condition at $z_{1}$) and $C_{2} = \sin^{-1}{\left(\sqrt{I_0/I_{\mathrm{max}}} \exp{\left(  \frac{\alpha L}{2}  \right)}\right)}$. The maximum intensity $I_{\mathrm{max}}$ is obtained for $z_{3} = z_{2} + \frac{2 C_{2}}{\alpha \tan(C_{2})}$.
\vspace{4pt}
\newline
The spatial spectrum associated with the target profile can be derived analytically using Eq.~\eqref{Spectrum}. As all the parts composing the target profile can either be expressed by an exponential rising function or a sine square function, computing the spatial spectrum associated to the following generic functions is sufficient: $I_{\mathrm{sin}}^{(i, j)}(z) = I \, \sin^{2} \left( a_{i} \, \frac{z-z_{i}}{z_{j}-z_{i}}+b \right)$ and $I_{\mathrm{exp}}^{(i, j)}(z) = I \, \exp \left[\alpha (z-z_{i}) \right)]$. The derivation of the associated spectra $S_{\mathrm{sin}}^{(i, j)}$ and $S_{\mathrm{exp}}^{(i, j)}$ is straightforward; we only give the final result :  
\begin{align}
\begin{split}
        S_{\mathrm{sin}}^{(i, j)} = {}& \sqrt{I} \, \frac{l}{k_{z}}  \left[a_{i} \, \frac{\cos(a_{i}) - \cos(a_{i}+b_{j})}{a_{i}^{2}-(\delta k l)^{2}} \right. \\ 
        & \hspace{-0.2cm} - \left. i \delta k \frac{\sin(a_{i}) \, e^{\,i \delta k z_{i}} - \sin(a_{i}+b_{j}) \, e^{\,i \delta k z_{j}}}{a_{i}^{2}-(\delta k l)^{2}} \right],
\end{split}
\end{align}
\vspace{-15pt}
\begin{equation}
        S_{\mathrm{exp}}^{(i, j)} = - \sqrt{I} \, \frac{2}{k_{z}} \frac{e^{\,i \delta k z_{i}}- \exp{\left(\alpha l/2\right)} \, e^{\,i \delta k z_{j}}}{\alpha+2i\delta k},
\end{equation}

\noindent where $l = z_{j}-z_{i}$ and $\delta k = k_{z0}-k_{z}$. We finally obtain the spectrum summing the spectral contributions coming from the different parts of the profile: $S = S_{\mathrm{sin}}^{(0, 1)} + S_{\mathrm{exp}}^{(1, 2)} + S_{\mathrm{sin}}^{(2, 3)} + S_{\mathrm{sin}}^{(3, 4)}$. 

\subsection{Clearing of the refractive stretching}

\vspace{-10pt}

 Let's assume that the target Bessel beam enters at $z_{1}$ a material with an attenuation coefficient $\alpha$ and a refractive index $n$. In order to counteract the refractive stretching of the beam, let's also replace $L$ with $L/n$ and $\alpha$ with $\alpha \, n$ in the second line of Eq.~\eqref{Spectrum12}. Using Eq.~\eqref{Spectrum} and the change of variable $z  \rightarrow  \Tilde{z} = n(z  -  z_{1})$, we can derive the spectrum $S_{1,2}$ associated to the exponential rising part of the on-axis profile (between $z_{1}$ and $z_{2}$):
\begin{align}
\label{Spectrum12}
    S_{1,2} &=  \sqrt{I_{0}} \, \frac{e^{\, i(k_{z0}-k_{z}) \, z_{1}}}{n \, k_{z}}  \int_{0}^{L} \exp{\left( \frac{\alpha \Tilde{z}}{2} \right)}  \, e^{\, i(k_{z0}-k_{z})\frac{\Tilde{z}}{n}} \, \mathrm{d}\Tilde{z} \nonumber \\
     & =  - \frac{i}{{n \, k_{z}}}\frac{\sqrt{I_{0}} \, e^{\,i(k_{z0}-k_{z}) \, z_{1}}}{\left(\frac{k_{z}-k_{z0}}{n}\right)  +  i\frac{\alpha}{2}}  \left(1-  e^{\,-i\left[\left(\frac{k_{z}-k_{z0}}{n}\right)+ i \frac{\alpha}{2}\right]L} \right) 
\end{align}
\noindent The on-axis electric field $E(r=0,z)$ is related to the spatial spectrum $S$ by the Fourier transform Eq.~\eqref{OnAxisField}. In practice, $k_{\mathrm{min}}$ and $k_{0}$ set respectively the lower and upper bounds of the integral coming in this equation (in air). Using Eq.~\eqref{Spectrum12} and Eq.~\eqref{OnAxisField} and the change of variable $\Bar{k}_{z}  =  (k_{z}  -  k_{z0})/n$, we can derive the on-axis electric field $E_{1,2} (r=0,z)$ associated to $S_{1,2}$ :
\begin{multline}
\label{OnAxisExp}
     E_{1,2}(r=0, \delta z)  =  \sqrt{I_{0}} \, e^{\,i \, k_{z0} (z_{1}+\delta z/n)} \\ \times \left[ \frac{-i}{\pi} \int_{0}^{\infty} \frac{1-e^{\,-i\left[\Bar{k_{z}}+ i \frac{\alpha}{2}\right]L}}{\Bar{k}_{z} +  i\frac{\alpha}{2}} e^{\,i \Bar{k}_{z} \delta z} \, \mathrm{d}\Bar{k}_{z} \right].
\end{multline}
\noindent As $z$ lies in the interval $[z_{1},z_{2}]$ and $z_{2} = z_{1} +L/n$, $\delta z  =  n(z-z_{1})$ varies from 0 to $L$. The phase $\Phi_{l}  =  k_{z0} \, (z_{1}  +  \delta z/n)$ is the phase accumulated by the Bessel beam along its propagation until $z$. The medium is supposed to be linear; this phase term is therefore the only expected. The inside brackets Eq.~\eqref{OnAxisExp} should then be real. Let's divide the integral in two parts $I_{1}$ and $I_{2}$ as follow :
\begin{align}
         I_{1}(\delta z)  &=  \frac{-i}{\pi} \int_{0}^{\infty} \frac{\Bar{k}_{z}  -  i\frac{\alpha}{2}}{\Bar{k}_{z}^{2}  +  \left(\frac{\alpha}{2}\right)^{2}} \, e^{\,i \Bar{k}_{z} \delta z} \, \mathrm{d}\Bar{k}_{z} \label{I1} \\ \label{I2}
         I_{2}(\delta z)  &=  \frac{i}{\pi} \, \exp{\left(\frac{\alpha L}{2}\right)} \int_{0}^{\infty} \frac{\Bar{k}_{z}  -  i\frac{\alpha}{2}}{\Bar{k}_{z}^{2}  +  \left(\frac{\alpha}{2}\right)^{2}} \, e^{\,-i \Bar{k}_{z} (L-\delta z)} \, \mathrm{d}\Bar{k}_{z}  
\end{align}
\noindent From Eq.~\eqref{I1} and Eq.~\eqref{I2}, we derive the real parts of $I_{1}$ and $I_{2}$ : $\mathrm{Re}\left(I_{1}\right) = 0 $ and $\mathrm{Re}\left(I_{2}\right) = \exp{\left( \frac{\alpha \delta z}{2} \right)}$. The on-axis electric field $E_{1,2} (r=0,\delta z)$ is finally given by :
\begin{align}
\label{Spectrum12}
  \begin{split}
    E_{1,2}\,(r=0,\delta z) ={}& \sqrt{I_{0}} \, e^{\,i \, k_{z0} (z_{1}+\delta z/n)} \\
         & \hspace{1.8cm} \times \left[ \mathrm{Re}\left(I_{1}\right) + \mathrm{Re}\left(I_{2}\right) \right] 
    \nonumber     
\end{split}\\
\begin{split}
     \phantom{E_{1,2}\,(r=0,\delta z)} ={}& - \sqrt{I_{0}} \, e^{\,i \, k_{z0} (z_{1}+\delta z/n)} \exp{\left( \frac{\alpha \delta z}{2} \right)}.
\end{split}
\end{align}
\noindent By replacing $L$ with $L/n$ and $\alpha$ with $\alpha \, n$ in the expression of the target on-axis intensity profile Eq.~\eqref{SystemProfile}, we manage to overcome the refractive stretching of the Bessel beam and compensate for the good attenuation coefficient $\alpha$.  

\bibliography{sample}


\end{document}